\documentclass[prl,groupedaddress,showpacs,byrevtex,twocolumn]{revtex4}
\usepackage[]{amssymb}
\usepackage[]{graphicx}

\begin{document}

\bibliographystyle{apsrev}

\title{Low-Energy Scale Excitations in the Spectral Function of Organic
  Monolayer Systems}

\author{J. Ziroff}
\author{S. Hame}
\author{M. Kochler}
\author{A. Bendounan\footnote{present address: SOLEIL L'Orme des Merisiers, Saint-Aubin - BP 48, 91192 GIF-sur-YVETTE CEDEX FRANCE SOLEIL, Paris, France}}
\author{A. Sch\"oll}
\email[corresponding author. Email: ]{achim.schoell@physik.uni-wuerzburg.de}
\author{F. Reinert}
\affiliation{Universit\"at W\"urzburg, Experimentelle Physik~VII \&
  R\"ontgen Research Center for Complex Material Systems RCCM,
  Am Hubland, D-97074 W\"urzburg, Germany}
\affiliation{Karlsruher Institut f\"ur Technologie KIT, Gemeinschaftslabor f\"ur
Nanoanalytik, D-76021 Karlsruhe, Germany}

\date{\today}

\begin{abstract}
Using high-resolution photoemission spectroscopy we demonstrate that
the  electronic structure of several organic monolayer systems, in particular 1,4,5,8-naphthalene
tetracarboxylic dianhydride and Copper-phtalocyanine on Ag(111), is
characterized by a peculiar excitation feature right at the Fermi level. This
feature displays a strong temperature dependence and is immediatly
connected to the binding energy of the molecular states, determined by the
coupling between the molecule and the substrate. At low temperatures,
the line-width of this feature, appearing on top of the partly
occupied lowest unoccupied molecular orbital of
the free molecule, amounts to only $\approx 25$~meV, representing an unusually
small energy scale for electronic excitations in these systems. We discuss
possible origins, related e.g. to many-body excitations in the
organic-metal adsorbate system, in particular a generalized Kondo
scenario based on the single impurity Anderson model.
\end{abstract}
\maketitle

For more than twenty years there has been a thorough investigation
on $\pi$-conjugated organic molecules,
which have shown to be suitable for the application in organic
electronic devices \cite{tang1986,schwoerer_book,umbach02}. These molecules
often form long-range ordered films on single-crystalline metal
substrates, allowing a systematic and fundamental study by various
surface sensitive techniques. In particular from
electron spectroscopy methods, as photoemission spectroscopy (PES),
inverse photoemission (IPES), and x-ray absorption (XAS), deep insight into many important features of the electronic
properties of condensed films and their interfaces has been achieved \cite{umbach1996,seki1997,hill2000,ueno2008,fahlmann2008}. The latter are of
particular importance since they crucially determine the properties of
possible devices. Additional 
microscopic and spectroscopic information has been obtained by use
of scanning tunneling microscopy (STM), leading to complementary
information about the relation between geometrical and electronic
structure \cite{kilian08,temirov08,wang2009,tseng2010}.

Usually, even the spectra of highly ordered films show features with
a line-width of several hundreds of meV, mostly determined by
vibronic excitations within the adsorbed molecule
\cite{krause2008,ueno2008,scholz09}. Therefore, even experiments with
modern high-resolution photoemission spectrometers for VUV
photoemission (UPS) display only features which are about two orders
of magnitude larger than the most narrow peaks in other solid state
or surface systems \cite{huefner_vher}. Local spectroscopic
measurements by STM, on the other hand, show narrow features in the
tunneling conductivity measurements through a single molecule
\cite{temirov08,fernandez08,Mugarza2011,Choi2010,Perera2010}, which have been attributed to a
possible Kondo like process within the charge transport through the
adsorbed molecule. However, neither a direct evidence of a local
magnetic moment, necessary for the Kondo effect, nor an immediate
experimental observation of the spectral function does exist yet.

Here we report about a high-resolution photoemission study on two
different organic monolayer systems that display a new narrow peak in
the excitation spectra near the Fermi level, possibly related to
strong electronic correlations in the system. Since the photoemission
spectrum can be interpreted as the single particle spectral function
\begin{math}$A$^{<}_{\text{\overrightarrow{k}}}$(E)$\end{math} \cite{huefner_vher,reinert05}, it allows in general a quantitative determination of the
influence of many-body effects in the system. After describing the experimental setup and the sample preparation, we
discuss 1,4,5,8-naphthalene tetracarboxylic dianhydride (NTCDA) on Ag(111), and compare it to Copper-phtalocyanine (CuPc) then,
which has an analogous spectral behaviour, although the geometrical
structure of the overlayer is different.

The experiments have been performed on a UHV setup based on a
high-resolution photoemission analyser (Gammadata R4000) in
combination with a monochromatized, microwave driven VUV source
(He~I: $h\nu\!=\!21.22$~eV, He~II: 40.8~eV). The sample temperature
during the measurements was varied between room temperature (RT) and
approximately 20~K (see Ref.~\onlinecite{klein08_cecu6} for
details). Substrate preparation (sputtering and annealing
\cite{cuagau_reinert01}), surface characterization by low-energy
electron diffraction (LEED) and x-ray photoemission spectroscopy (XPS), and
deposition of the molecules from a Knudsen cell \cite{ziroff09} have
been done at RT and {\em in situ}. Due to radiation damage, the
spectra have shown a decrease of the LUMO (lowest unoccupied molecular orbital of
the free molecule) intensity by about 20\%
after an exposition time of $\approx30$~min to the VUV light.
Therefore, we have minimized the exposition time to less than 5~min
per spectrum, repeatedly repositioned the sample to non-irradiated
areas, and carefully checked the reproducibility of the spectra.

The phase diagram of NTCDA/Ag(111) shows several phases in
dependence on temperature and coverage. Here we restrict ourselves
on the so-called {\em relaxed} phase, which is characterized by a
commensurate superstructure explained in detail in \cite{stahl98,schoelljcp2004,schoell10} representing a net coverage of $\Theta\,=\,0.7$ with respect to  a dense monolayer \cite{stahl98}. The preparation produces highly ordered
monolayers, leading to clear LEED patterns and,
consequently, to a well defined backfolding of the Ag~5sp bulk bands
in the ARUPS data \cite{hame_dipl,ziroff10}.

\begin{figure}[t]
  \begin{center}
    \includegraphics[width = 8cm]{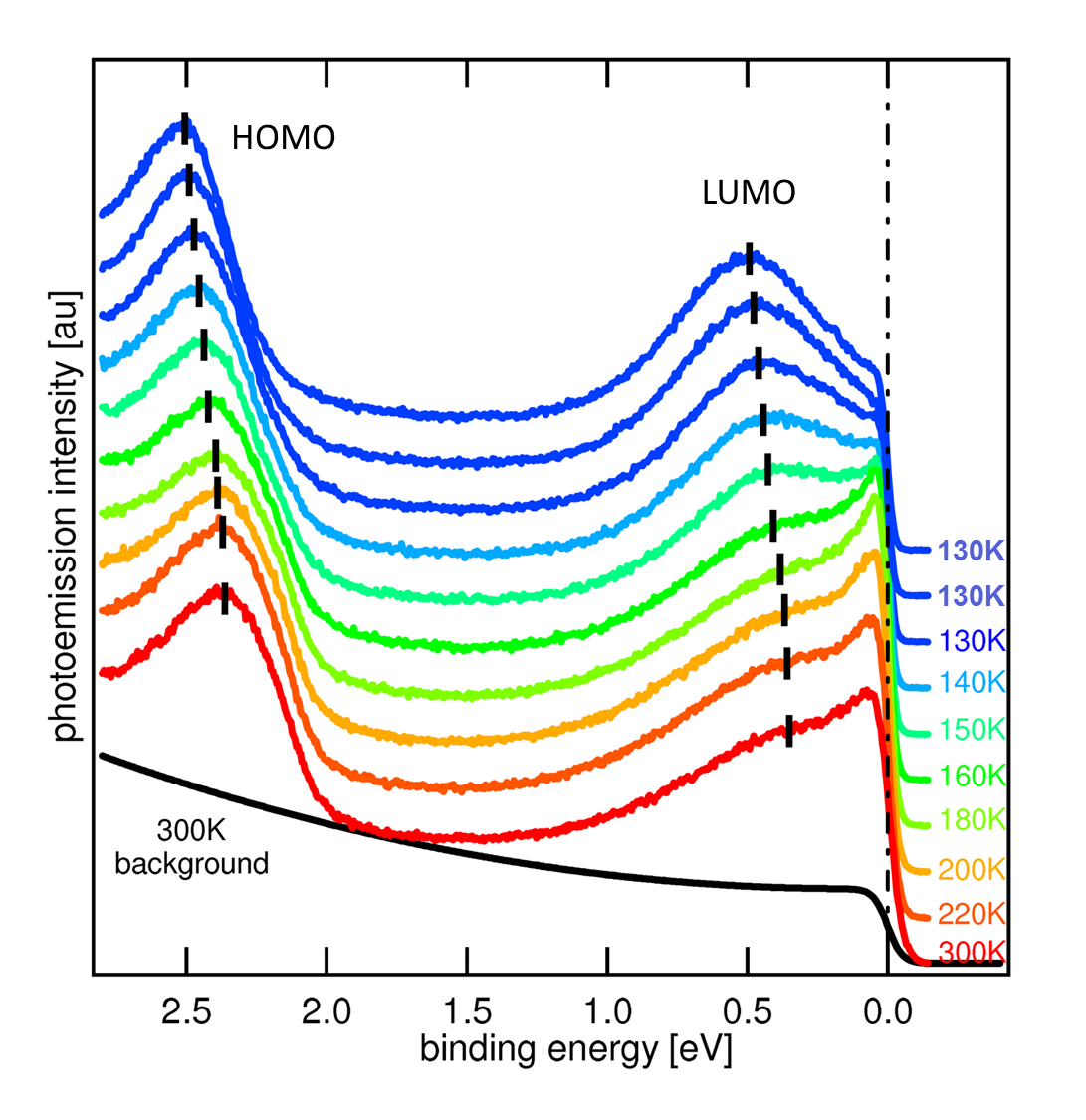} 
    \caption[tempseries]{(Color online) Series of subsequent high-resolution photoemission spectra
      (He{\sc I}$_\alpha$, emission angle \begin{math}$35$^{o}\end{math} ) on the relaxed monolayer of NTCDA/Ag(111) for different temperatures. The cooling was
      performed slow (over a period of 3~h) leading to an order-disorder phase transition \cite{schoell10}. The top three spectra were recorded subsequently at a constant temperature of 130 K.  
The solid black line at the bottom indicates the
contribution of the substrate background.
    \label{fig:tempseries}}
  \end{center}
\end{figure}

It was already demonstrated elsewhere \cite{bendounan07} that the
occupied valence regime of the monolayer system is dominated by
the two molecular states HOMO (highest occupied molecular orbital) and LUMO or
hybridization state, in reference to the electronic orbitals of the
free NTCDA molecule. These states appear above the ``background'' of
the metal substrate states, and show a characteristic angle
dependence of the photoemission intensities \cite{ziroff10} (see supplement),
whereas an energy dispersion in dependence on $k_\|$ can not be
observed \cite{bendounan07}. As in the related system PTCDA/Ag(111)
\cite{zou06,kilian08,ziroff10} the LUMO, unoccupied in case of the
isolated molecule, is cut by the Fermi edge, demonstrating a {\em
partial} occupation of this state by a transfer of electrons from
the substrate to the molecule. This is equivalent with a significant
hybridization of the LUMO with occupied states of the
metal substrate. Note that the LUMO does only appear in the
photoemission spectra of the molecules in the first layer. For molecules farther away
from the interface the LUMO remains unoccupied \cite{bendounan07}.

The strong interaction between the chemisorbed molecules and the
substrate leads to a large shift of the Shockley-type surface state,
which is characteristic for the (111) faces of noble metals
\cite{cuagau_reinert01}, to energies above the Fermi
level \cite{schwalb08,sachs09,schwalb10}. In case of NTCDA/Ag(111) the Shockley state is found 400 meV above $E_F$ \cite{Marks2011}.

\begin{figure}[t]
  \begin{center}
    \includegraphics[width = 9cm]{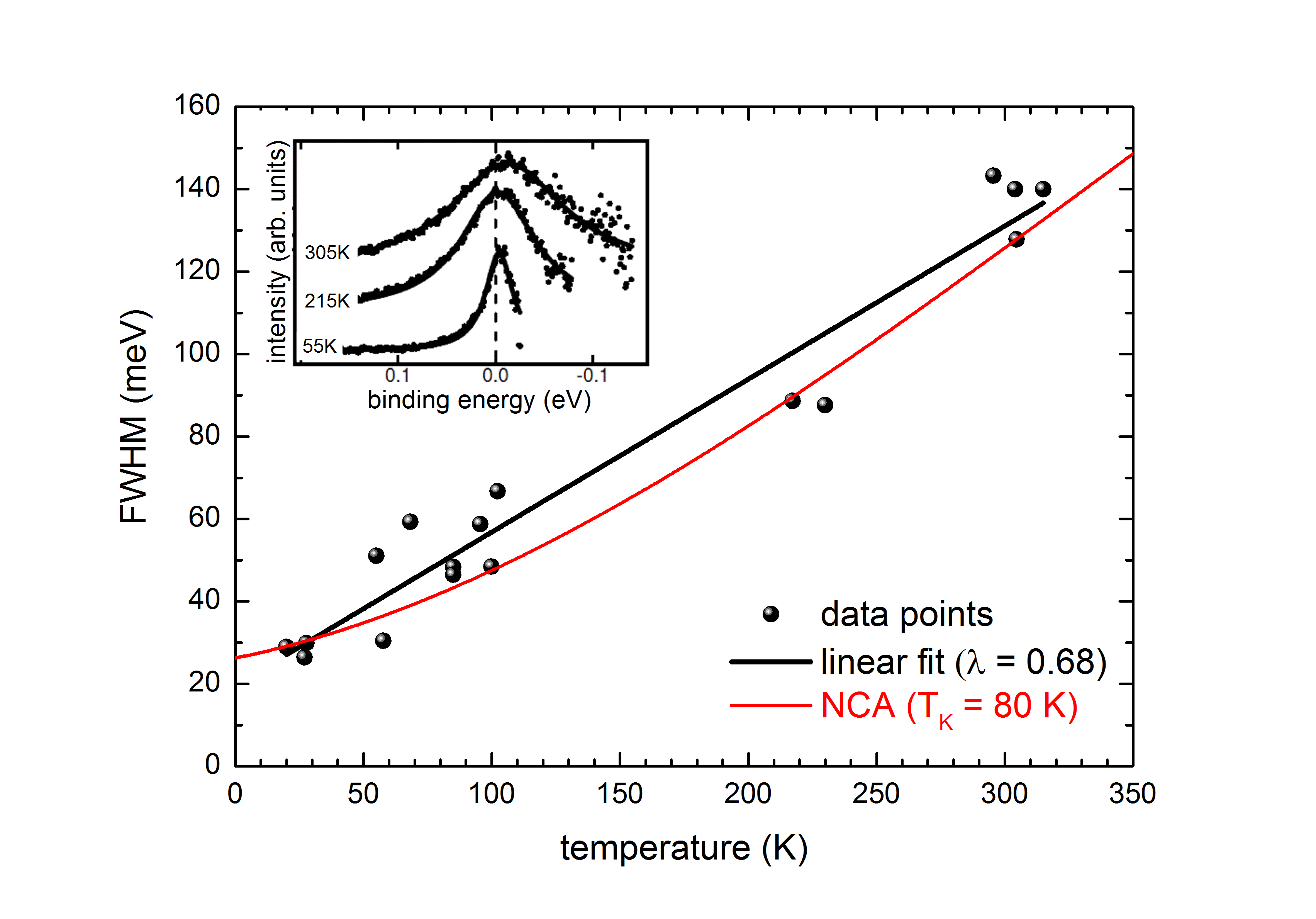}
    \caption[tempdep]{(Color online) Analysis of the PES linewidth (FWHM) of the
      narrow peak at $E_F$ in case of NTCDA/Ag(111). The spectra were divided by the FDD, background corrected and
      fitted by a Lorentzian afterwards (see examples in the inset,
      normalized to maximum height). The resulting line widths 
      are given vs. the sample temperature $T$. Assuming an electron-phonon broadening mechanismus, the
      linear behavior can be described by a electron-phonon coupling
      parameter of $\lambda\!=\!0.68$ (black line). The red (dark gray) curve was derive from a NCA modelling with $T_K$~=~80~K \cite{costi-1996}.
    \label{fig:tempdep}}
  \end{center}
\end{figure}

Fig.~\ref{fig:tempseries} shows a series of high-resolution, normal
emission PES spectra of 1~ML NTCDA/Ag(111) for
different temperatures. The intensity maxima of the two molecular
features closest to the Fermi level, HOMO and LUMO, appear at
approximately 2.4~eV and 0.4~eV, respectively. The typical line-width of
these molecular states is of the order of 500~meV in monolayer
systems. The background of the substrate is flat near the Fermi
level and rises steeply at a binding energy of $\gtrsim4$~eV.
Already in the room temperature (RT) spectrum, the most peculiar
feature is an additional narrow peak right at the Fermi level,
appearing on top of the comparatively broad LUMO. This new peak can
be observed for all emission angles and shows the same angle
dependence in its intensity as the LUMO (see supplement), which next to the energy shift described above \cite{Marks2011} is
another clear evidence that it is not related to the Shockley-state
appearing around the $\bar{\Gamma}$-points of the reconstructed
surface Brillouin zones \cite{forster04} but a molecular feature. Furthermore, it can not be
explained by the molecular states of the free NTCDA molecules alone.
Therefore, it is reasonable to suppose that this state is a
consequence of the close interplay between Bloch states of the
substrate and molecular states at the interface.

Very important for the understanding of the new narrow feature is
the analysis of its temperature dependence.
Fig.~\ref{fig:tempseries} shows a series of normal emission spectra
for temperatures between room temperature (bottom) and $T\!=\!130$~K
(top three spectra). The sample was cooled slowly in this case, i.e.
the time between first (RT) and last spectrum (130~K) amounts to
3~h. Note that the influence of contamination can be ruled out by core level PES \cite{schoelljcp2004} and X-ray absorption data \cite{schoell10}. The spectra display several characteristic changes with decreasing temperature: 1.) HOMO and LUMO shift to higher binding
energies by about 130~meV, 2.) the new peak at the Fermi level
becomes narrower and 3.) looses rapidly in intensity at temperatures
below 170~K until the peak has nearly completely vanished at 130~K.
Keeping the temperature at constant 130~K, one can observe further
changes with the same trend over several hours. Going back to RT,
the initial spectrum will be restored again (not shown).

The described temperature dependence of the spectra is immediately
related to an order-disorder transition \cite{schoell10}. Below 180~K the long-range order
within the NTCDA monolayer is distroyed by a thermally activated rearrangement of the molecules. The stable low-temperature phase is
amorphous, which can be seen by the disappearance of both the LEED
spots and the backfolding of the substrate bands in the
photoemission data. However, since this order-disorder transition is
connected to a thermally activated rearrangement of the molecules,
the transition can be prevented if one cools down rapidly, leading
to a ``frozen'' phase with the long-range ordered superstructure of
the relaxed phase at RT. This frozen phase can be stabilized for several hours, which
allows for a detailed study of the intrinsic line-width of the
narrow peak in the ordered phase even down to low temperatures.

Fig.~\ref{fig:tempdep} shows the result of a quantitative line width
analysis. The individual values for the full width at half maximum
(FWHM) were obtained by a two step analysis: To restore the
photoemission signal above $E_F$, the spectra have been
normalized to the Fermi-Dirac distribution (FDD)
\cite{klein08_cecu6,greber97,kondo_reinert01,reinert_lecturenotes07,ehm07} first, and then fitted by a
Lorentzian, which describes the line shape of the peak reasonably
well (see inset of Fig.~\ref{fig:tempdep}). Within the experimental
errors, the narrow peaks appear exactly symmetrically to $E_F$,
independent from the temperature.

The temperature dependence of the line width
follows roughly a linear behavior, down to minimum values of about 25~meV
at the lowest temperatures accessible in our experiment. This value contains the energy resolution and contributions from extrinsic broadening effects as discussed in \cite{cuagau_reinert01}. Such a linear temperature dependence is known from the high-temperature limit
of the Debye model, where the slope is
immediately related to the electron-phonon coupling parameter $\lambda$
by $\partial \Gamma/\partial T = 2\pi\lambda k_B$ \cite{reinert_lecturenotes07}. From a linear
least-squares fit of the temperature dependence we obtain here
$\lambda\!\approx\!0.7$, which would indicate a
very strong electron-phonon coupling. This is not far from characteristic values of
typical superconducting metals ($\lambda\!\approx\!1$)
\cite{reinert_pb03} or C60-systems \cite{knupfer93,gunnarsson1995,yang2003} where also a
characteristic satellite structure was observed. The coupling parameter within the metal substrate, on the
other hand, is only of the order of $\lambda\!\approx\!0.1$
\cite{eiguren02}. Particularly for the similar PTCDA/Ag(111), electron energy
loss spectroscopy \cite{tautz02} indicates substantially smaller coupling values than what one can derive from the slope in fig.~\ref{fig:tempdep}. Obviously a strong electron-phonon coupling alone can
not describe the observed peculiar spectral feature in our photoemission data. 
Note that the temperature dependence of the intensity would bear additional information but can not be evaluated in the present case since the peak is directly at $E_F$ and the normalisation to the FDD, which is crucial for the analysis of the line shape, can lead to artefacts in intensity \cite{ehm07}.

\begin{figure}[t]
  \begin{center}
    \includegraphics[width = 8cm]{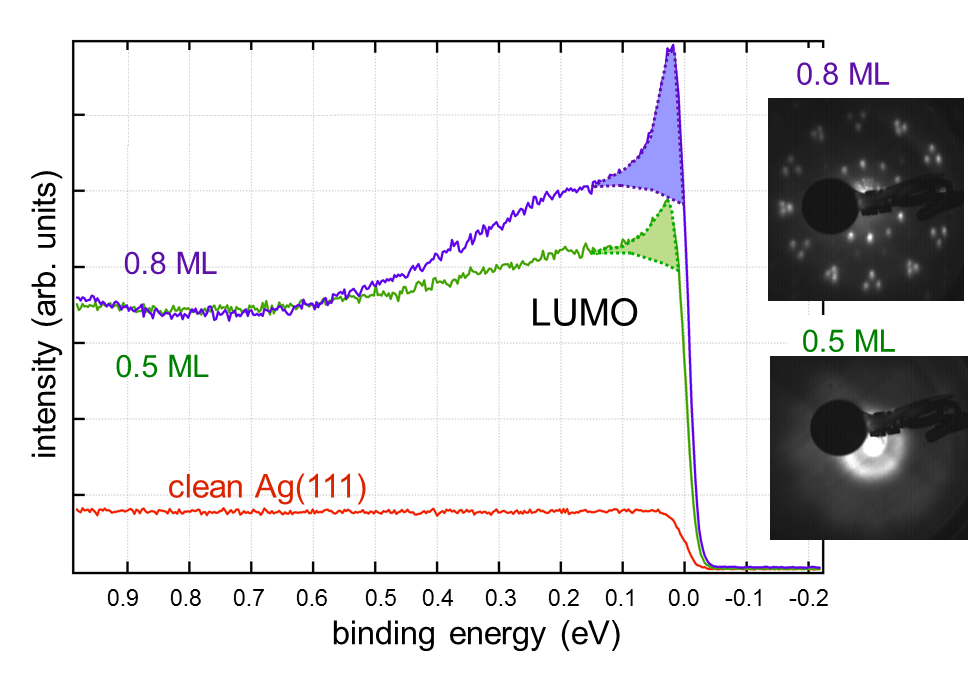}
    \caption[CuPc] {(Color online) He~{\sc I$_\alpha$} spectra (emission angle  \begin{math}$50$^{o}\end{math}) of Cu-phthalocyanine
(CuPc) for two different coverages $\Theta\,=\,0.5$ (green (light gray)) and $\Theta\,=\,0.8$ (blue (dark gray)) at $T\,=\,80$~K. For low coverages, there
is no long range order formed, as displayed by the LEED
patterns ($E_0\,=\,12$~eV). However, there is a resonance peak at $E_F$.
    \label{fig:CuPc}}
  \end{center}
\end{figure}

To clarify the importance of the long-range lateral order,
we compare the results of NTCDA/Ag(111) to the data of CuPc submonolayers on Ag(111),
which show many similarities. At constant temperature this system
forms long-range ordered phases for high coverages, whereas the
individual molecules remain separated and disordered for low
coverages \cite{kroeger10, stadtmueller11}. The respective spectra
at temperatures of $T\,=\,80$~K are shown in Fig.~\ref{fig:CuPc}. For
$\Theta\,=\,0.5$ (green spectrum on bottom) the LEED-picture in the inset shows the
disordered g-phase, while for $\Theta\,=\,0.8$ (blue spectrum on top) the ordered c-phase is established. Note that the
molecules do not form islands and the surface is thus completely
covered in both cases. The overall photoemission intensity from the
molecules is of course approximately a factor of two smaller than in the high
coverage spectrum when normalized to the Ag background, simply
because the number of molecules per area is smaller. Apart from
that, the two spectra look identical: as in the case of
NTCDA/Ag(111),  the former LUMO is partly occupied and appears below
the Fermi level ($E_B\,\approx\,0.13$~eV). Most important, for both
samples the narrow peak at the Fermi level is clearly observed. The
width (FWHM) of this signal is FWHM
$\approx 40$~meV and is the same for both coverages. Again, the line width is strongly temperature dependent and increases following a linear behavior with $\lambda\approx\!0.9$ when interpreted as due to electron-phonon coupling.

Therefore it is evident, that long range lateral order of the adsorbate molecules is not necessary for the existence of the narrow peak. Instead,
it is the position of the LUMO which is important for the
development of this additional feature. In the case of
NTCDA/Ag(111), there occurs a temperature and time dependent shift
of the LUMO to higher binding energies, immediately connected to
the disappearance of the narrow peak.

With other words, the exact position of the LUMO is determined by
hybridisation between the Bloch states of the substrate and the
localized molecular states of the individual molecule. Such a
scenario is described by the very versatile {\em single impurity
Anderson model} (SIAM), which is based on the interplay between a
localized magnetic impurity, e.g. a transition metal or lanthanoid
atom, with the conduction states of a metal \cite{anderson61}. The SIAM has been applied
successfully for the explanation of the physical properties of many
transition metal and rare-earth compounds \cite{gschneidner}, and
even for the quantitative estimation of many-body effects in
adsorbate systems (Newns-Anderson) \cite{newns69}. In particular, by
use of numerical techniques \cite{gunnarsson83L,bickers87} one can
calculate the spectral function of such an impurity system, giving
the so called {\em Kondo resonance} at the Fermi level. The maximum
position and the width of the Kondo resonance reflect the small
energy scale $k_BT_K$ of the system, which determines all electronic
and magnetic low-energy excitations in the system. Note that this
model is based on a single impurity, coherence effects and the
formation of an impurity lattice are not included and therefore not
necessary for the appearance of the Kondo resonance. A typical temperature dependence for the SIAM line width derived from a non-crossing approximation (NCA) calculation \cite{costi-1996} is shown in fig.~\ref{fig:tempdep} ($T_K$=80~K, off set by an additional broadening of 12~meV). A possible quantitative influence from coupling to phonons \cite{Mugarza2011,fernandez08} is not discussed here. 

Indeed, the spectroscopic parallels between typical Kondo (actually
Heavy Fermion) systems, as e.g. CeCu$_2$Si$_2$
\cite{kondo_reinert01} or CeCu$_6$ \cite{klein08_cecu6} are
striking: a narrow feature appears at the Fermi level with a strong
temperature dependence in both line width and intensity.

A quantitative determination of the Kondo scale $k_BT_K$ from the temperature dependence requires a
more detailed knowledge of the model parameters Coulomb
correlation energy $U$, single particle energy $\epsilon_f$ and
hybridization strength $V$. However, a comparison with the spectra
of inorganic systems allows a rough estimation of $k_BT_K$. For
example, the Kondo temperature must be significantly larger than for
CeCu$_6$ ($T_K\,\approx\,5$~K), where the Kondo resonance vanishes
nearly completely in the raw data for temperatures above 100~K. From
the  comparison with other Ce compounds, particularly CeSi$_2$
\cite{ehm07}, we estimate $T_K$ to be on the scale of roughly 100~K. This is in accordance with the low-T limit of the experimental line width (25~meV), which represents an upper limit of $k_BT_K$, and with the NCA simulation in fig.2.  

The interpretation in the frame of the SIAM has further important
implications, namely the meaning of the position of the LUMO. As
shown in Fig.~\ref{fig:tempseries}, the narrow feature disappears, when the
binding energy of the LUMO increases. Indeed, the investigation of
other related organic adsorbate systems
\cite{bendounan07,kilian08,ziroff09,scholz09,stadtmueller11} shows
that only if the LUMO has its maximum below but close to the Fermi
level --- this might be seen as slightly above half filling --- the
narrow feature appears. This importance is known from the 2D Hubbard
model \cite{dagotto92,preuss95} with a band filling in the range of 0.6--0.7.

Finally, for a ``Kondo like'' ground state, the correlation energy
$U$ must be large in comparison to the other model parameters, in
particular to the hybridization between conduction band and
localized states $V$ and to the single particle binding energy
$\epsilon$, possibly given by the position of the LUMO maximum in
this case, which is about 0.3~eV.  An upper limit of 2~eV for U can be derived from the change in HOMO-LUMO separation from non-interacting molecules in multilayers to NTCDA at the interface. In addition, for the similar, yet larger, molecule PTCDA $U$ is estimated to about 400~meV if adsorbed on Ag(111) \cite{greuling2011}. Following simple size arguments this value can be regarded as a lower limit for $U$ of the smaller NTCDA. Moreover, a rough estimate of 0.2~eV can  be derived for $V$ from the additional broadening that occurs for the LUMO in comparison to the HOMO (which shows weaker hybridisation \cite{ziroff10}).

In conclusion, we could demonstrate that photoemission excitation
spectra of chemisorbed organic molecules show a narrow feature at
the Fermi level with a strong characteristic temperature dependence,
that can not be explained by a mere electron-phonon coupling
process. The appearance of the feature is strongly dependent on the
position of the LUMO, i.e. on the band filling of the system, but
does not depend on the long-range order of the monolayer. Therefore,
we conclude that the observed spectral resonance at the Fermi level
is a result of the coupling between the individual molecule and the
conduction electrons of the metallic substrate. For an
interpretation within the SIAM, the model parameters can be
estimated to reasonable values. However, further quantitative
studies by numerical methods are required to understand the
many-body properties of these molecular systems.

\begin{acknowledgments}
We would like to thank Hans Kroha (University of Bonn) for helpful discussions.
This work was supported by the Deutsche Forschungsgemeinschaft
(FOR1162 and GRK 1221) and the Bundesministerium f\"ur Bildung und Forschung BMBF (grant no. 05K10WW2 and 3SF0356B). 
\end{acknowledgments}

\bibliography{sb,ep2,pes,kondo,organics,xtcda,huefner,others}

\end{document}